# Supporting Patients in Managing Electronic Health Records and Biospecimens Consent for Research: Insights from a Mixed-Methods Usability Evaluation of the iAGREE Portal

Di Hu, MS[1], Xi Lu, PhD[2], Yunan Chen, PhD[1], Michelle Keller, PhD[3], An T. Nguyen, OTD[4], Vu Le, MS[1], Tsung-Ting Kuo, PhD[5], Lucila Ohno-Machado, MD, PhD[5], Kai Zheng, PhD[1]
[1]University of California, Irvine, Irvine, CA, USA; [2]University at Buffalo, Amherst, NY, USA; [3]University of Southern California, Los Angeles, CA, USA; [4]Cedars-Sinai Medical Center, Los Angeles, CA, USA; [5]Yale University, New Haven, CT, USA

**Abstract**

*De-identified health data are frequently used in research. As AI advances heighten the risk of re-identification, it is important to respond to concerns about transparency, data privacy, and patient preferences. However, few practical and user-friendly solutions exist. We developed iAGREE, a patient-centered electronic consent management portal that allows patients to set granular preferences for sharing electronic health records and biospecimens with researchers. To refine the iAGREE portal, we conducted a mixed-methods usability evaluation with 40 participants from three U.S. health systems. Our results show that the portal received highly positive usability feedback. Moreover, participants identified areas for improvement, suggested actionable enhancements, and proposed additional features to better support informed granular consent while reducing patient burden. Insights from this study may inform further improvements to iAGREE and provide practical guidance for designing patient-centered consent management tools.*

**Introduction**

De-identified electronic health records (EHRs) and biospecimens can be shared and used for research without explicit patient consent under the Health Insurance Portability and Accountability Act (HIPAA) privacy rule[1]. These data have been traditionally considered low risk and are therefore used in research without a requirement for patient consent. However, achieving complete anonymity is not feasible for health data[2], and the risk of re-identification has long been recognized[3]. This risk has escalated with recent advancements in AI, as sophisticated algorithms can analyze large datasets, uncover patterns, link de-identified data with different data sources, and potentially identify individuals[4]. Lack of patient awareness and control can raise ethical concerns about privacy and autonomy[5].

Public awareness of data privacy, security, and ownership has increased, along with the demand for greater transparency and control over health data sharing[6,7]. Research shows that patients want to be informed and provide consent for the use of their health data, whether identifiable or not[5,8,9]. Studies further indicate that patients prefer granular control over which data items they share and with whom they share them[9–13], potentially withholding sensitive information or restricting access to certain organizations. When offered transparency and granular control, patients express increased trust and a stronger willingness to contribute their health data for research[9,12,13]. Given this trend, data use without granular consent may, in the future, discourage patient support and continuous participation in health research[6,11]. Therefore, we decided to study whether granular patient control over sharing EHRs and biospecimens, even when de-identified, is technically feasible and how patients can be better supported in expressing their preference for data sharing.

The Office of the National Coordinator for Health Information Technology has advocated for granular consent choices and the transition from paper-based to interoperable, electronic, and auditable consent processes[14]. However, few solutions have been successfully implemented in real-world settings. In our previous pilot study, we developed and deployed iCONCUR, a tiered electronic informed consent system for research[9]. Over 1,200 participants used this system to provide consent, specify sharing preferences, and receive notifications about their data use during the study period. While iCONCUR was found to be viable, usability challenges remained, particularly regarding ease of navigation. Hence, we developed iAGREE, an enhanced portal that builds on iCONCUR and incorporates design considerations from a focus group study with patient and researcher stakeholders[13]. In this usability study, we adopted a user-centered, mixed-methods approach to evaluate and refine iAGREE, aiming for a user-friendly design that can effectively support patients in managing consent for the research use of their de-identified EHRs and biospecimens. Through two rounds of portal usability assessments, we identified areas for improvement and patient suggestions on enhancing the portal design to facilitate informed consent management while minimizing burden.

Insights from our study can guide further improvements to the iAGREE and inform the design of similar tools to empower patients in managing data sharing for research.

**Method**
We developed the iAGREE portal, informed by our previous studies[9,13], and conducted two rounds of usability assessments. Feedback from the first round was used to refine the design, and the second round was conducted to identify any remaining usability issues. A mixed-methods approach was used to analyze data collected from the usability assessment. This study was approved by the Institutional Review Board of the University of California, Irvine (UCI).

*iAGREE Portal Design and Development*
The iAGREE portal is a patient-centered platform that enables patients to manage granular consent for sharing de-identified EHRs and biospecimens for research. The portal has three main functions:

1. Default Preference Setup: Patients can set default sharing preferences for various data types (e.g., demographics, mental health information) across nonprofit, for-profit, and government research organizations. These default preferences can be updated anytime by the patient, and the system will prepopulate consent choices for new data requests based on these preferences. However, patients can still modify their choices when reviewing individual studies.
2. Review Data Requests: iAGREE allows users to review specific data requests and either approve or decline them individually or apply their pre-set default preferences. Each data request shows the requesting organization, study title, and a brief study description. If patients want more details, they can review the full consent form before making a decision. The system enables patients to selectively share certain data types while withholding others for a given study request.
3. Review and Modify Consent History: Patients can track their previous consents and withdraw their consent for all or specific data types, provided the data has not yet been shared. Once data is shared, it cannot be withdrawn, but patients can prevent future use by modifying their preferences.

Additionally, iAGREE allows patients to explain why they chose not to share certain data for specific studies. The interface is designed for both desktop and mobile screens. For selected screenshots, see Figure 1.

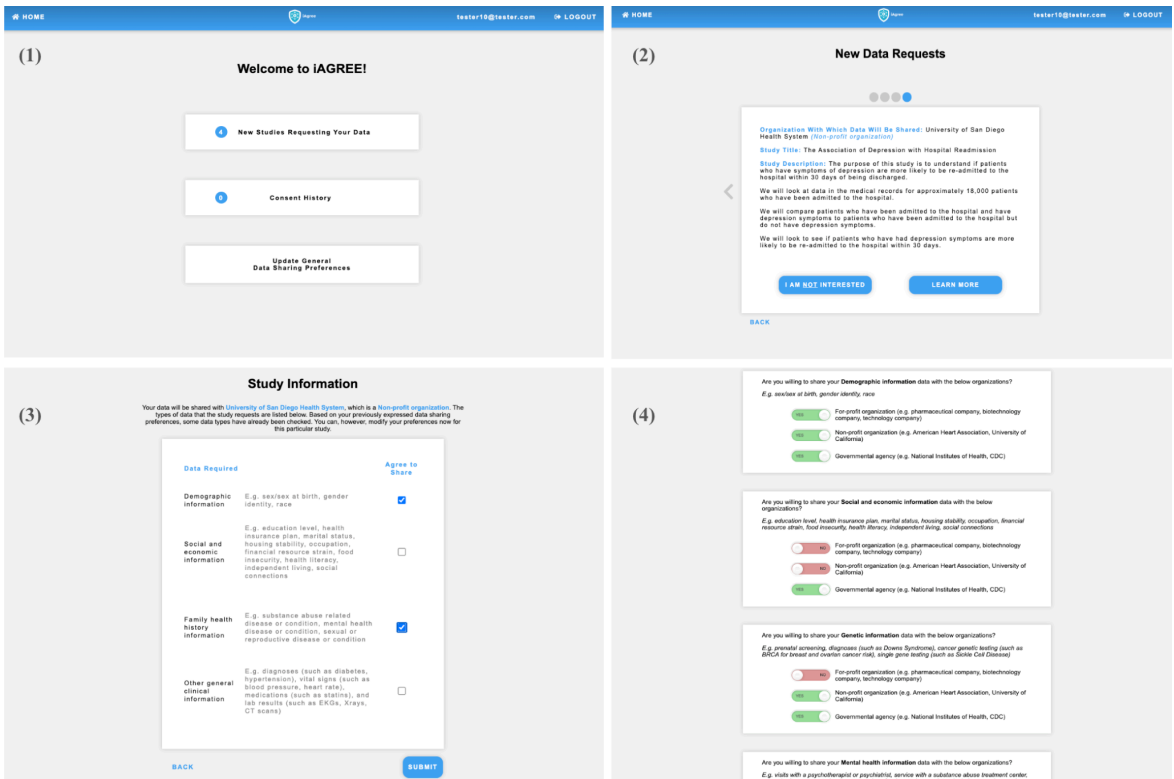

**Figure 1.** (1) iAGREE home page; (2) new data request page; (3) review/modify and submit data sharing preferences for an individual study; (4) default preferences setup page.

Following the first round of usability assessment, we refined the portal design based on participant feedback. Updates primarily focused on clarifying the iAGREE portal's purpose and features as well as enhancing the clarity of instructional language. These updates were prioritized based on their frequency in participant feedback and the feasibility of implementation within the short timeframe between usability assessments. Complex improvements, such as multilingual support, were deferred for future implementation. The iAGREE portal was developed using ASP.NET, along with Vanilla JavaScript and a Microsoft SQL Server database.

*Participant Recruitment*

A total of 40 participants were recruited from three health systems in Southern California: UCI Health, the University of California, San Diego (UCSD) Health, and Cedars-Sinai Medical Center. Each round of usability assessment included 20 participants. Eligibility criteria required participants to be at least 18 years old, fluent in English, and actively receiving care at one of the participating health systems. Participants were recruited through study flyers in outpatient clinics, emails sent by research coordinators, social media pages of the participating health systems, and outreach to the Patient and Family Advisory Board. Contact information of the research team were provided, and interested participants reached out to the research team via email or text message. Usability assessment sessions were conducted either in person in a private meeting room at UCI or remotely via Zoom, depending on participant preference. Given the minimal risk posed by the research, a waiver of signed consent was obtained, and verbal consent was collected prior to the study session. Each participant received a $25 Amazon gift card as compensation for their time and contributions. Either a phone number or email address was collected solely for gift card distribution and was immediately deleted after the gift card was delivered. No other identifiers were collected in this study.

*Usability Assessment Session and Data Collection*

The usability assessment session followed a structured format designed to evaluate multiple aspects of usability, including ease of navigation, clarity of information, and efficiency in completing tasks. The process remained the same across two rounds of the usability assessments. Participants completed the following steps to evaluate the iAGREE portal's usability:

1. Task-Based Usability Testing: Participants interacted with the portal to complete five predefined tasks simulating common user actions: (a) setting up default or general data-sharing preferences; (b) accepting a data-sharing request; (c) declining a data-sharing request; (d) modifying data-sharing decisions for a previously approved/declined request; and (e) updating default data-sharing preferences.
2. Think-Aloud[15]: Participants were asked to verbalize their thoughts while interacting with the portal, allowing researchers to capture real-time usability questions, concerns, and cognitive processes. This method provided insights into how users perceive and interpret different elements of the interface.
3. In-Session Interview: Follow-up questions were asked as needed when participants shared usability-related perceptions during the task-based interaction. At the end of the portal interaction, a brief semi-structured interview was conducted to capture participants' overall experiences, additional reflections, and suggestions for improvement. For example, interview questions included "*Is there anything you would like to share or add about your experience using iAGREE?*" and "*Do you have any [other] suggestions for improving the iAGREE to make it easier to use and understand?*"
4. Post-Session Survey: After completing the portal interaction and interview, participants filled out a post-session survey that included demographic questions and the 19-item Post-Study System Usability Questionnaire (PSSUQ)[16] a validated instrument for assessing software usability. Participants rated PSSUQ items on a scale from 1 (strongly agree) to 7 (strongly disagree), with a "Not Applicable" option, providing quantitative measures of satisfaction and system effectiveness.

All sessions, including both in-person and virtual, were audio- and screen-recorded using Zoom. Each session lasted up to one hour, and audio recordings were transcribed. Observational notes from the screen recordings were incorporated into the transcripts for analysis. The post-session survey was built and managed using REDCap (Research Electronic Data Capture). All identifiable information was removed from the transcripts before analysis.

*Data Analysis*

After the first round of usability assessments with 20 participants, the first author analyzed the transcripts and screen recordings, focusing on observed usability issues and participant suggestions for solutions. The research team identified a list of feasible improvements, which were then implemented in the iAGREE portal. Qualitative data from both rounds of usability assessments were analyzed using thematic analysis[17]. The first author familiarized herself with the data and open-coded 10 transcripts to generate initial codes. Two authors (DH and KZ) reviewed and discussed these codes to identify key patterns in participants' usability feedback on the iAGREE portal. Following this initial review, the first author re-coded the initial 10 transcripts and coded the remaining ones, focusing on participants' suggestions for portal design to support patients in sharing their EHRs and biospecimens for research. Similar codes were grouped into themes and iteratively reviewed, compared, and refined through team discussions. Qualitative coding was conducted using ATLAS.ti (ATLAS.ti GmbH, Berlin, Germany). The post-session survey results were analyzed quantitatively using descriptive statistics to summarize participant characteristics and assess PSSUQ responses. All descriptive analyses were performed in R (version 4.4.1).

**Results**

In this section, we present findings from our mixed-method evaluation. We first describe participant characteristics, then report the PSSUQ results, quantifying the usability of the current iAGREE portal. Finally, we highlight key themes from qualitative participant feedback.

*Participant Demographics*

Of the 40 participants, 20 were recruited from UCI Health (1st round: n = 8, 2nd round: n = 12), 10 from UCSD Health (1st round: n = 5, 2nd round: n = 5), and 10 from Cedars-Sinai (1st round: n = 7, 2nd round: n = 3). As shown in Table 1, the majority of our participants were between 20 and 39 years old at the time of the study and identified as female, non-Hispanic, and Asian. Most participants accessed the iAGREE portal using a laptop or desktop computer while others used a mobile phone or tablet.

**Table 1.** Participants demographics and their devices used for accessing the iAGREE portal.

| Demographics | All (n = 40) | 1st Round (n = 20) | 2nd Round (n = 20) |
|---|---|---|---|
| **Age Group n (%)** | | | |
| 20 - 29 | 21 (52.5) | 13 (65) | 8 (40) |
| 30 - 39 | 13 (32.5) | 2 (10) | 11 (55) |
| 40 - 49 | 0 (0) | 0 (0) | 0 (0) |
| 50 - 59 | 1 (2.5) | 1 (5) | 0 (0) |
| 60 + | 5 (12.5) | 4 (20) | 1 (5) |
| **Gender, n (%)** | | | |
| Female | 24 (60) | 14 (70) | 10 (50) |
| Male | 16 (40) | 6 (30) | 10 (50) |
| **Race, n (%)** | | | |
| Asian | 24 (60) | 9 (45) | 15 (75) |
| Black or African American | 5 (12.5) | 3 (15) | 2 (10) |
| White | 7 (17.5) | 5 (25) | 2 (10) |
| More than one race | 2 (5) | 2 (10) | 0 (0) |
| Other | 1 (2.5) | 1 (5) | 0 (0) |
| Prefer not to answer | 1 (2.5) | 0 (0) | 1 (5) |
| **Ethnicity, n (%)** | | | |
| Hispanic or Latino | 2 (5) | 1 (5) | 1 (5) |
| Not Hispanic or Latino | 37 (92.5) | 18 (90) | 19 (95) |
| Prefer not to answer | 1 (2.5) | 1 (5) | 0 (0) |
| **Device, n (%)** | | | |
| Desktop or laptop | 31 (77.5) | 16 (80) | 15 (75) |
| Tablet | 2 (5) | 1 (5) | 1 (5) |
| Phone | 7 (17.5) | 3 (15) | 4 (20) |

*PSSUQ Results*

All participants completed the PSSUQ. Since none encountered error messages while using the iAGREE portal, they either selected "Not Applicable" or "Neither Agree nor Disagree" for the two PSSUQ items related to error messages (item 9) and error recovery (item 10). As a result, these two items were excluded from the final analysis. PSSUQ scores range from 1 to 7, with **lower scores indicating higher satisfaction**. The mean overall satisfaction score, calculated by averaging scores across all items, was 2.17 ± 1.40 for all participants, reflecting a generally positive perception of the iAGREE portal's usability. The second-round evaluation yielded a lower mean overall satisfaction score with less variability (1.81 ± 0.74) compared to the first round (2.52 ± 1.79), indicating an improved satisfaction. Details about other item groups related to system usefulness (item 1-8), information quality (item 9-15), interface quality (item 16-18) are reported in Table 2.

**Table 2.** Mean scores for PSSUQ item groups by participant groups.

| **Mean Scores (Standard Deviation)** | **All** | **1st Round** | **2nd Round** |
|---|---|---|---|
| Overall Satisfaction | 2.17 (1.40) | 2.52 (1.79) | 1.81 (0.74) |
| System Usefulness | 1.95 (1.51) | 2.41 (2.00) | 1.50 (0.52) |
| Information Quality | 2.20 (1.43) | 2.64 (1.79) | 1.76 (0.74) |
| Interface Quality | 2.63 (1.74) | 2.63 (1.83) | 2.62 (1.70) |

Figure 2 shows the mean scores for individual PSSUQ items, comparing first-round, second-round, and overall participant responses. Overall, the item with the lowest mean score was item 2, indicating strong agreement that the portal was easy to use. Conversely, the item with the highest mean score was item 18, suggesting that some participants felt the portal lacked certain expected functionalities. For first-round participants, item 7 ("*It was easy to learn to use iAGREE*") received the most agreement, while item 18 was rated the worst. In the second round, item 3 ("*I could effectively complete the tasks using iAGREE*") had the most favorable rating, whereas item 16 ("*The interface of iAGREE is pleasant*") received the least. Notably, most items had lower mean scores in the second round, indicating overall improved usability after portal refinements. However, two interface-related items, item 16 and item 17 ("*I liked using the interface of iAGREE*"), had relatively higher scores in the second round. The most substantial differences between the two rounds were observed in item 11, asking if participants agree that the information provided with this system is clear.

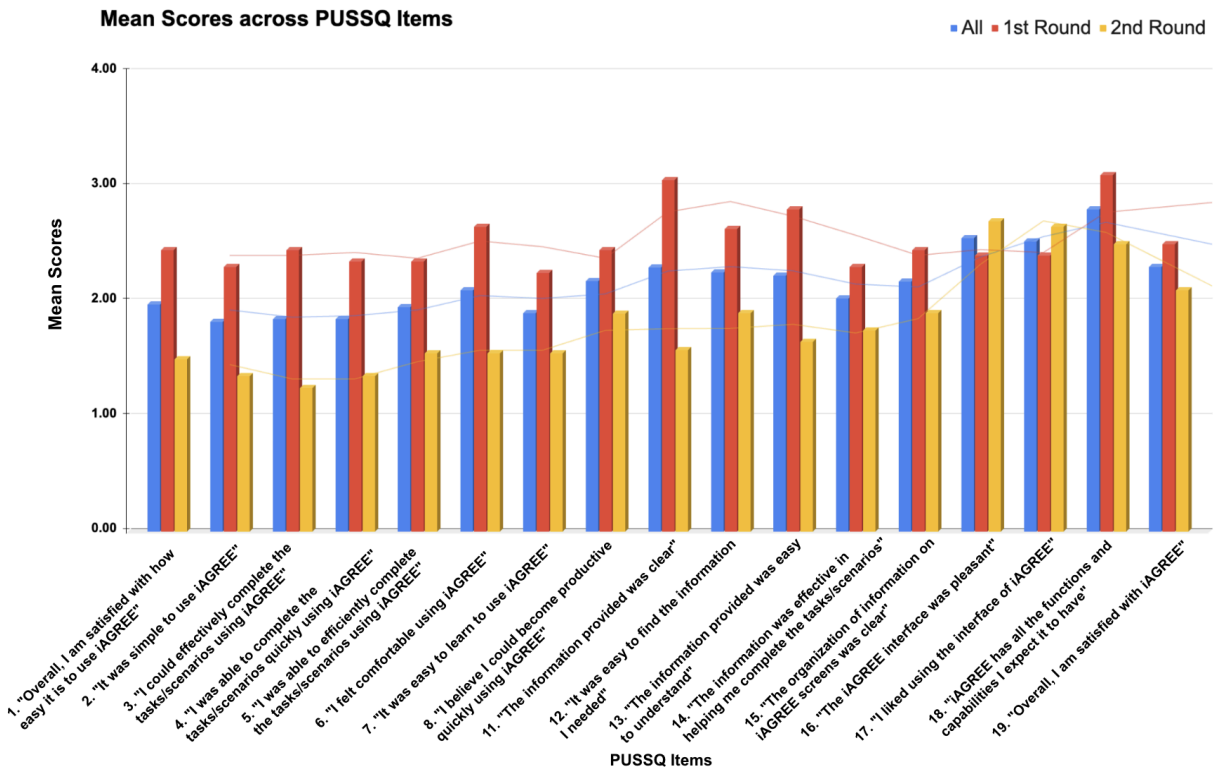

**Figure 2.** Mean scores for individual PSSUQ items, compared across all, first-round, and second-round responses.

*Qualitative User Feedback and Suggestions for Improvement*
Overall, participants found the iAGREE portal easy to navigate, with nearly all providing positive feedback on its usability. Most described the user interface as "*clear*" or "*straightforward*", making it "*simple*" to set up data-sharing preferences or respond to data requests (n = 28). Beyond usability feedback, participants identified areas for improvement, suggested specific enhancements, and proposed additional features to better assist patients in managing granular consent for sharing EHRs and biospecimens for research. Five key themes emerged, reflecting participants' suggestions that the portal should: (1) clarify its purpose and functionality, (2) add detailed information for data-sharing decisions, (3) enhance support for diverse comprehension needs, (4) optimize consent management workflows to reduce patient burden, and (5) enable communication with data recipients (researchers).

*Clarify the Purpose and Functionality of the Portal.* Participants found the iAGREE portal well-structured and easy to use but noted that it could more clearly communicate its purpose and functionality. Although iAGREE was introduced at the beginning of usability assessment sessions, some participants still found it unclear that the portal was designed specifically for **sharing de-identified data** for **research**. They suggested adding explicit statements to ensure patients understand that iAGREE is not meant for other purposes, such as patient care or marketing. Some also recommended reiterating key messages, such as "*Your identity will not be revealed,*" (P2) on relevant portal pages to reassure patients. Additionally, participants emphasized that the portal needs to specify whether consent via iAGREE applies solely to **sharing existing EHRs and specimens** or if it also requires participation in new data collection activities. A few asked clarifying questions, such as "*If I am interested in sharing, does it mean that I need to participate, or do they just need my data?*" (P16) These questions were particularly common when reviewing biospecimen-sharing preferences, as some participants were unsure whether agreeing to share meant they would need to provide new samples. While participants appreciated the portal's features for managing granular consent and modifying consent history, **the limited presence of such features in current consent practices led some to question how they work and what practical impact they might have.** Several raised questions about whether partial data sharing is possible and whether consent decisions could be modified over time. For example, P24 asked, "*If I opt out of a requested [data] category such as mental health information, would this essentially not share any*

*information?*" while P14 inquired, "*Can you modify [a previous consent] at any time, or will it change at some point when the study starts?*" To avoid these confusions, participants suggested refining the interface with **clearer descriptions and strategically placed explanations to help users fully understand its purposes and functionalities.** Explanations about iAGREE's purpose and consent history modification were incorporated into the portal refinements after the first-round usability assessment, leading to fewer questions on these aspects in the second round.

*Add Detailed Information for Data-Sharing Decisions.* Participants valued that iAGREE provided request descriptions and study information, but some felt that the information was not sufficiently comprehensive or detailed. Interestingly, even when participants understood that their identifiable information (e.g., name, birth date, medical record number) would not be requested for sharing, they still sought **more detailed explanations about data privacy and security**. For example, P26 considered a general confidentiality statement inadequate and wondered how data recipients would "*guarantee the de-identification.*" while P27 asked about procedures in case of a data breach. Other participants raised concerns about the potential for their de-identified data to be misused or sold, with P13 questioning "*Can they [researchers or institutions] sell my data? Even if it is de-identified, what is the legality of that?*" Similarly, although the portal explained that shared data would contribute to research, participants expressed a desire for more concrete and relatable examples of how data sharing could directly benefit them or other patients, advance scientific knowledge (P34), or support medical education (P23). Additionally, a few participants were unsure why certain data types, such as socioeconomic information, were being requested and stated that they needed clearer justifications for their relevance to research. Participants suggested expanding existing content by incorporating **more detailed security explanations, real-world examples of research impact, and clear rationales for each requested data type.** These additions would help patients weigh relevant risks and benefits and make more informed data-sharing decisions.

*Enhance Support for Diverse Comprehension Needs.* While participants desired additional details for decision-making, they acknowledged that the volume and complexity of information could be overwhelming, emphasizing the need to support patients with diverse comprehension abilities and preferences. They highlighted that the portal's current strategies, such as bolding key text and color coding, were helpful but suggested **more effective information presentation and interface design** to further facilitate comprehension. Many emphasized the importance of using clear, explicit language and avoiding abbreviations. P12 noted that instructions and terminology should be tailored to the general public and those with lower literacy levels, while several others recommended incorporating tooltips to provide plain-language explanations when medical jargon was unavoidable. For lengthy and dense text, such as study request details, participants suggested making the portal **more interactive and utilizing more visual aids**. They envisioned an interface that strategically used positioning, highlighting, and pop-ups to help users quickly identify the critical information that matters most. P40 specified the need to grasp "*the most important information of a study*" while P29 wanted "*the information I care the most*" to stand out. Participants also noted that the portal could be more inclusive. A few suggested adding **multilingual support** for non-English speakers, while others recommended **accessibility features** to better accommodate individuals with disabilities. Many of the suggestions for clearer and more readable language were integrated after the first round of usability testing, resulting in fewer comprehension challenges in the second round.

*Optimize Consent Management Workflows to Reduce Patient Burden.* Participants appreciated the granular control iAGREE provided over their sharing preferences for individual data requests, yet they recognized potential burdens if the request list became long. Many praised the portal's ability to automatically apply default preferences to new requests but suggested **further optimization of the consent process by minimizing unnecessary or repetitive actions.** Some proposed adding a "*Select All*" option to consent to multiple studies simultaneously (P6, P28) or an option to approve all requests from a specific type of organization (P4). Another well-received feature was the inclusion of "Not Applicable" response options, which allowed participants to bypass irrelevant questions. For example, some participants mentioned that they did not or would never have pregnancy-related information, so having a "Not Applicable" option for this data type avoided unnecessary responses. Additionally, due to the scope of this usability study, the evaluated iAGREE portal was not integrated with participants' EHRs. As a result, some participants noted that the simulated requests they received were not always relevant to them and, therefore, emphasized the importance of implementing a "*pre-screening*" process to ensure that data requests are only sent to "*patients who qualify [the study inclusion criteria]*" (P31) so that they are not "*seeing studies that aren't relevant*" (P35). Finally, to improve **long-term usability,** participants recommended adding status labels to remind them of their previous consent choices without needing to click into each study. Some also suggested adding sorting and search functions to improve request and consent management. These changes would reduce the effort required to manage consents over time, making the process more efficient.

*Enable Communication with Data Recipients (Researchers)*: While the current iAGREE application allows patients to leave feedback, some participants wanted the ability to **more actively communicate with researchers** requesting their data. Several appreciated that the portal provided a space for them to explain to researchers why they declined sharing certain data for mutual understanding. P10 noted, "*I am doing similar [consent] forms to this. It's frustrating to leave questions unanswered… it's nice to be asked.*" However, some expressed a need to ask researchers specific questions about studies in a more timely and direct manner before deciding whether to share their data. Others were interested in tracking study progress and receiving updates on research outcomes. P37 and P39, for example, hoped to follow up on the study progress, while P35 believed it would be encouraging to have "*a mechanism to find out what happened, learn the results, or see [if] a paper came out of it.*" Participants suggested three ways to enable more direct and active communication: **including researchers' contact information in study descriptions, adding a real-time study progress board, or incorporating a built-in chat feature within the portal.** By introducing a structured way for patients to connect with researchers, iAGREE could foster greater transparency and engagement in the research process. Participants saw this connection not only a means to obtain additional information, but also as a way to understand how their contributions made a meaningful impact.

**Discussion**

Our study not only evaluated and refined the iAGREE portal but also uncovered deeper insights into patient preferences regarding the sharing of de-identified health data for research. Overall, both qualitative and quantitative user feedback indicated that participants were satisfied with the portal's usability. More importantly, the findings highlight practical, patient-informed design considerations for developing a user-friendly portal that facilitates an informed, engaged, and low-burden granular consent process. Furthermore, results from two rounds of PSSUQ assessments demonstrate that user satisfaction, although initially high, can further improve following refinements based on participant suggestions for clarifying portal information. These improvements were reflected in participants' higher perceived information quality and system usefulness. This finding underscores the critical role of information clarity in supporting patient consent[18] and highlights the effectiveness of an iterative, user-centered approach to evaluating and designing electronic consent platforms[5,19]. The only aspect that did not show notable improvement in the second round of PSSUQ results was the interface quality score. However, cross-checking with qualitative feedback suggests this may be due to fewer usability issues affecting navigation and comprehension. With functional and comprehension concerns largely addressed, participants likely shifted their focus to aesthetic preferences, such as color choices, layout structure, and additional visual elements to create a lively user interface.

While participant feedback primarily focused on the usability of the iAGREE portal, it naturally expanded to reveal broader implications for understanding and supporting patient preferences in managing consent for de-identified health data sharing for research. We found that when participants praised portal features—such as dedicated sections for setting preferences by data type and organization, the ability to review individual data requests, and the option to provide partial data—they were, in fact, reinforcing their preference for granular control over data sharing. This preference has been consistently observed in numerous studies on patient perspectives regarding data control[9–13]. Our portal design successfully accommodates this preference, and our study findings indicate ways to further enhance support through patient-suggested portal improvements. Additionally, our findings confirm that even when sharing de-identified data, participants continue to seek transparency, remain concerned about data privacy and security, and engage in a similar decision-making process as they do when sharing protected health information[5,9,20]. However, managing granular consent requires patients to review individual data requests in detail, which increases cognitive burden and makes it difficult to comprehend excessive or complex information when making informed decisions. This tension between control and usability emerged in our findings and has also been observed in other electronic consent studies[5,19,21]. To ease this tension, our participants proposed specific and actionable next steps for improving consent platform design. Designers should more strategically utilize interactive features and visual aids to facilitate efficient information comprehension for informed consent. Our findings also suggest that system workflows should be streamlined to minimize redundant user actions, reducing the burdens on patients who may already feel overwhelmed by managing consent at an individual level. Personalization of information presentation and consent processes can be implemented to accommodate patients' varied information and control needs while ensuring they retain access to sufficient details and granular control.

Another key finding that can inform the future design of consent platforms is patients' interest in maintaining a connection with researchers during and after sharing their data. Across multiple studies, patients have expressed a strong desire to learn about progress or outcomes of research using their data[5,9,13]. However, these studies have not explored how patients prefer to access this information. Our participants offered specific suggestions, such as a built-in chat, researcher contact information, or a progress board—allowing them to voluntarily reach out or check for updates as needed. This approach contrasts with the notification-based approach assumed in previous research[9]. Additionally, Morse et al. found that researchers also hope to connect with patients to educate them about the value of their data and the consequences of unrepresentative samples[13]. Therefore, incorporating features that facilitate mutual communication between patients and researchers may foster trust and strengthen patient engagement in research. Moreover, dedicated efforts beyond the consent platform environment should be made to develop patient education resources that support informed consent and promote data sharing. According to the iCONCUR pilot, participants from a site with educational programs on health and digital consent were more willing to share their data for research[13]. Our participants admitted their limited understanding of how non-health data (e.g. social and economic factors) support research and how research by for-profit organizations benefits them or society. They mentioned that real-world examples illustrating benefits of these research could serve as strong motivators for data sharing. Therefore, educational resources should not only explain common medical terms and EHR concepts[22], but also clarify what it means to share health data for research, including potential risks and benefits. Eventually, these resources can be integrated into consent management platforms to empower patients, enhance their participation in research, and ultimately benefit both patients and researchers.

Our study provides valuable insights into the design and development of electronic consent management tools, but several limitations should be considered. First, our sample consisted of participants recruited from three health systems in Southern California, which may not fully represent the perspectives of patients from other regions or healthcare settings. Additionally, although our sampling approach spanned a wide age range, the majority of our participants were under 40 years old. This younger, potentially more digitally literate population may not reflect the needs and experiences of older adults, who are often participants in health research and may face different usability challenges. Future studies should examine the generalizability of our findings across more diverse populations, including those with varying ages, levels of digital literacy, and familiarity with health data. Second, because our design refinements between the two rounds of usability assessments were intended to further identify potential areas for improvement, we implemented only selective updates. Future iterations of iAGREE should consider incorporating additional participant suggestions, such as multilingual support and accessibility enhancements, to ensure the portal is inclusive. Third, while iAGREE includes functionality for updating patient preferences and modifying consent, long-term user engagement with the system remains unexplored. It is essential to assess whether and how patients continue to manage their data sharing over time and to explore how such a portal can effectively support them in a longitudinal, dynamic, real-world setting. Finally, this study focused on usability evaluation rather than regulatory or operational considerations related to implementation. In future work, we plan to investigate the integration of patient-centered consent models within broader healthcare frameworks, addressing challenges such as interoperability with existing EHR systems and researcher-facing data request and delivery mechanisms.

**Conclusion**
This study successfully evaluated and refined the iAGREE portal, a patient-centered electronic consent platform that enables patients to exercise granular control over the sharing of de-identified EHRs and biospecimens for research. Our findings provide empirical insights into the usability of the iAGREE portal and highlight key patient-informed suggestions for improving the design of such platforms. Future designs should incorporate these suggestions to develop user-friendly electronic consent management systems that facilitate an informed, engaged, and low-burden granular consent process. Additionally, our findings reaffirm patients' desire for transparency and control over their health data, even when de-identified. Platforms like iAGREE can contribute to a future in which patients are actively engaged in a more transparent and ethical research ecosystem, ultimately promoting the responsible use of EHRs and biospecimens in research for the benefit of all.

**Funding and Acknowledgement**
This study was funded by NIH/National Human Genome Research Institute (NHGRI) grant number R01HG011066 (PI: Ohno-Machado). The authors acknowledge the research participants for their time and contributions.